\documentclass[12pt]{article}
\usepackage[margin=2.5cm]{geometry}
\usepackage{subcaption}
\usepackage{mathtools}

\usepackage{graphics} 
\usepackage{epsfig} 
\usepackage{amsmath} 
\usepackage{amssymb}  
\usepackage{url}

\usepackage{amsmath}
\usepackage{color}
\usepackage{footnote}
\usepackage{algorithm}
\usepackage{algpseudocode}
\usepackage{algorithmicx}
\usepackage{multirow} 
\usepackage{booktabs}
\usepackage{graphicx}
\usepackage{amssymb}
\usepackage{amsbsy}
\usepackage{array}
\usepackage{longtable}
\usepackage{epstopdf}
\usepackage{pbox}
\usepackage{breqn}
\usepackage{mathrsfs}
\usepackage{multicol}
\usepackage{supertabular}
\usepackage{enumerate}
\usepackage[colorinlistoftodos]{todonotes}
\usepackage{bbm}
\usepackage{bbold}
\usepackage{url}
\usepackage[justification=centering]{caption}
\newtheorem{thm}{Theorem}

\newtheorem{prp}{Proposition}

\title{\LARGE \bf
Security Risk Analysis of the Shorter-Queue Routing Policy for Two Symmetric Servers
}

\author{Yu Tang, Yining Wen, and Li Jin
\thanks{This work was in part supported by NYU Tandon School of Engineering and the C2SMART University Transportation Center.}
\thanks{Y. Tang and L. Jin are with the Department of Civil and Urban Engineering and C2SMART University Transportation Center, and Y. Wen is with the Department of Mechanical and Aerospace Engineering, New York University Tandon School of Engineering, Brooklyn, NY, USA. (emails: tangyu@nyu.edu, lijin@nyu.edu, yw3997@nyu.edu)}
}

\begin{document}
\maketitle
\thispagestyle{plain}
\pagestyle{plain}

\begin{abstract}
In this article, we study the classical shortest queue problem under the influence of malicious attacks, which is relevant to a variety of engineering system including transportation, manufacturing, and communications.
We consider a homogeneous Poisson arrival process of jobs and two parallel exponential servers with symmetric service rates.
A system operator route incoming jobs to the shorter queue; if the queues are equal, the job is routed randomly.
A malicious attacker is able to intercept the operator's routing instruction and overwrite it with a randomly generated one.
The operator is able to defend individual jobs to ensure correct routing.
Both attacking and defending induce technological costs.
The attacker's (resp. operator's) decision is the probability of attacking (resp. defending) the routing of each job.
We first quantify the queuing cost for given strategy profiles by deriving a theoretical upper bound for the cost.
Then, we formulate a non-zero-sum attacker-defender game, characterize the equilibria in multiple regimes, and quantify the security risk. We find that the attacker's best strategy is either to attack all jobs or not to attack, and the defender's strategy is strongly influenced by the arrival rate of jobs.  
Finally, as a benchmark, we compare the security risks of the feedback-controlled system to a corresponding open-loop system with Bernoulli routing. We show that the shorter-queue policy has a higher (resp. lower) security risk than the Bernoulli policy if the demand is lower (resp. higher) than the service rate of one server.
\end{abstract}

{\bf Keywords}:
Queuing systems, dynamic routing, attacker-defender game, security.

\section{Introduction}

The shorter-queue policy is a classical routing policy applicable to a variety of engineering systems, including transportation \cite{hung2015optimal}, production lines \cite{van2001performance}, and communications \cite{gupta2007analysis}. The idea of this routing policy is that a job is allocated to a server with a shorter queue when it arrives, which has been proved to be optimal if the system operator has perfect observation of the system states and perfect implementation of the policy \cite{ephremides1980simple}.
Such sensing and actuating typically rely on cyber components connected via wired or wireless communications. Although such connectivity can significantly improve throughput and reduce delay, it is vulnerable to malicious remote attacks and thus brings security risks.
In intelligent transportation systems, researchers have shown that traffic sensors and traffic lights can be easily intruded and manipulated \cite{ghena2014green,chen2018exposing}.
Similar security risks also exist in production lines \cite{lee2008cyber} and communication networks \cite{deng2002routing}.
However, such risk has not been well modeled and studied in the setting of queuing systems.

In this paper, we develop a game-theoretic model for the two-queue system subject to malicious attacks and estimate the security risk by characterizing the steady-state queue lengths and the game equilibrium.
We consider a homogeneous Poisson arrival process of jobs and two parallel exponential servers with symmetric service rates.
A system operator route incoming jobs to the shorter queue; if the queues are equal, the job is routed randomly.
A malicious attacker is able to intercept the operator's routing instruction and overwrite it with a randomly generated one.
The operator is able to defend individual jobs to ensure correct routing.
Both attack and defense induce technological costs.
The attacker's (resp. operator's) decision is the probability of attacking (resp. defending) the routing of each customer.
The attacker (resp. operator) is interested in maximizing (resp. minimizing) the long-time-average network-wide queuing cost minus the attacking cost (resp. plus the defending cost).

Numerous results have been developed for the two-queue system with perfect routing, i.e. perfect sensing plus perfect actuating \cite{flatto1977two,ephremides1980simple,halfin1985shortest,nelson1989approximation,foley2001join,eschenfeldt2018join}. Although some of these results provide hints for our problem, they do not directly apply to the security setting with imperfect sensing and/or actuating.
The two-queue system has been studied with delayed \cite{kuri1995optimal}, erroneous \cite{beutler1989routing}, or decentralized information \cite{ouyang2015signaling}, which provides insights for our purpose.
Based on previous results about the behavior of the generalized two-queue problem \cite{foley2001join}, we show that the two-queue system is stable in the face of attacks if and only if the probability of a successful attack, which is equal to the product of the probability of a job being attacked and the probability of a job not being defended, is less than the ratio between the service rate of one server and the jobs' arrival rate.
We further present an upper bound for the queue length, which we use as an approximation for queuing cost.

Next, we characterize the Nash equilibrium of the attacker-defender game.
Game theory is a powerful tool for security risks analysis that has been extensively used in various engineering systems \cite{manshaei2013game,wu2018securing,etesami2019dynamic}.
Game theoretic approaches have been applied to studying security of routing in transportation \cite{wu2018signaling,laszka2019detection} and communications \cite{bohacek2002enhancing,guo2016routing}.
However, to the best of our knowledge, the security risk of feedback routing policies has not been well studied from a perspective combining game theory and queuing theory, which is essential for capturing the interaction between the queuing dynamics and the players' decisions.
We quantitatively characterize the security risk (in terms of additional queuing cost and technological cost for defense) in various scenarios.
We show that the game has multiple regimes for equilibria dependent on the technological costs of attacking and of defending as well as the demand.
A key finding is that the attacker would either attack no jobs or attack all jobs.
When the attacking cost is high, the attacker may have no incentive to attack any jobs; consequently, the defender does not need to defend any jobs.
When the attacking cost is low, the attacker will attack every job; in this case, the defender's behavior will depend on the defending cost.
The regimes also depend on the arrival rate of jobs: for higher arrival rates, the attacker has a higher incentive to attack, and the defender has a higher incentive to defend.
In particular, if the arrival rate is less than the service rate of one server and the defense cost is prohibitively high, then the defender may has zero incentive to defend any jobs, regardless of the attacker's action.

As a benchmark for the closed-loop shorter-queue routing policy, we also consider an open-loop Bernoulli routing policy. The Bernoulli policy allocates each arriving job to either server with equal probabilities. In this case, the attacker can still intercept the routing instruction and replace it with a falsified one, and the attacker can protect a job from being attacked.
It is well known that the shorter-queue policy is in general more efficient than the Bernoulli policy in the nominal setting with perfect sensing and actuating.
In the presence of security failures, we show the shorter-queue policy has a higher (resp. lower) security risk than the Bernoulli policy if the demand is lower (resp. higher) than the service rate of one server.

The contributions of this paper are as follows.
First, we develop a formulation for security risk analysis of the two-queue system by synthesizing a queuing model and a game-theoretic model.
Second, we quantify the relation between the queuing cost and the actions for the attacker and the defender as well as key parameters of the model.
Third, we characterize the equilibria of the attacker-defender game and derive practical insights for dynamic routing.
Finally, we compare the security risks of the closed-loop shorter queue policy with the open-loop Bernoulli policy.
The rest of this paper is organized as follows.
In Section~\ref{sec_model}, we introduce the two-queue model and derive the queuing cost. In Section~\ref{sec_game}, we formulate the attacker-defender game and characterize the structure of the equilibria in various regimes.
In Section~\ref{sec_compare}, we compare the shorter-queue routing policy with the Bernoulli routing policy in terms of both nominal efficiency and security risk.
\section{Parallel queuing system facing attacks}
\label{sec_model}

\begin{figure}[htb]
\centering
\includegraphics[width=0.5\textwidth]{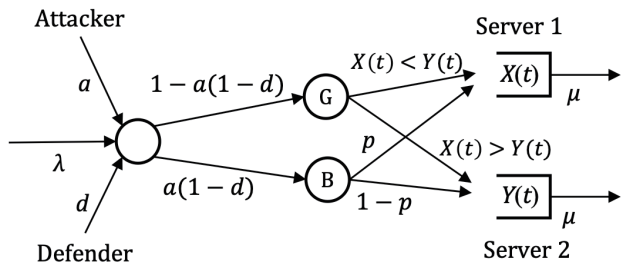}
\caption{Two-queue system with shorter-queue routing and malicious attacks.}
\label{fig_twoqueue}
\end{figure}

Consider the parallel queuing system in Fig.~\ref{fig_twoqueue}.
Jobs arrive according to a Poisson process of rate $\lambda$. Each server serves jobs at an exponential rate of $\mu$.
We use $X(t)$ and $Y(t)$ to denote the number of jobs, including waiting and being served, in the two servers, respectively.
The state space of the parallel queuing system is $\mathbb Z_{\ge0}^2$.

In the absence of attacks, the system operator has perfect observation of the states $X(t)$ and $Y(t)$. When a job arrives at time $t$, the operator allocates it to the shorter queue. That is, the job is allocated to server 1 (resp. 2) if $X(t)<Y(t)$ (resp. $X(t)>Y(t)$); if $X(t)=Y(t)$, then the job is allocated to each server with probability 1/2.

A malicious attacker is able to compromise the operator's dynamic routing. When a job arrives and is being allocated, the attacker is able to intercept the operator's routing instruction and replace it with a random one. Consequently, the job may be mistakenly allocated to the longer queue. Attacks have no impact when the queues are equal. Each job is attacked with probability $a\in[0,1]$, where $a$ is selected by the attacker.
When a job's routing is attacked, the original routing instruction is overwritten; instead, the job is routed to server 1 with probability $p$ and to server 2 with probability $1-p$, where $p\in[0,1]$ is selected by the attacker.
The system operator is able to protect a job's routing. When a job is protected, its routing is guaranteed to be correct, i.e. going to the shorter queue. The probability that a job is protected is $d\in[0,1]$, which is selected by the operator.
If the attacker does not attack or if the defender defends, which happens with probability $(1-a(1-d))$ a job goes to the good node (``G'' in Fig.~\ref{fig_twoqueue}) and is routed by the shorter-queue policy. Otherwise, the job goes to the bad node (``B'' in Fig.~\ref{fig_twoqueue}) and is routed randomly.

In the rest of this section, we first discuss the stability of the queues in the face of malicious attacks. Then, we provide a theoretical upper bound for the queue length, which we will use as an approximation of queuing cost.

\subsection{Stability}

It is well known that, in the absence of malicious attacks, the two-queue system is stable if and only if the demand is less than the total capacity, i.e. $\lambda<2\mu$.
In the result below, we show that malicious attacks can destabilize the queuing system.

\begin{prp}
The parallel queuing system is stable (i.e. positive recurrent) if and only if
\begin{subequations}
\begin{align}
    &\lambda<2\mu,\label{eq_stable1}\\
    &a(1-d)p\lambda<\mu,\\
    &a(1-d)(1-p)\lambda<\mu.\label{eq_stable3}
\end{align}
\end{subequations}
\label{prp_1}
\end{prp}
\noindent\emph{Proof}.
The queuing system is equivalent to a two-queue system with three classes of jobs. The first class enters server 1 as a Poisson process of rate $a(1-d)p\lambda$.
The second class enters server 2 as a Poisson process of rate $a(1-d)(1-p)\lambda$.
The third class arrive at the two-queue system as a Poisson process of rate $(1-a(1-d))\lambda$; when a job of this class arrives, the job joins the shorter queue. Thus, by \cite[Theorem 1]{foley2001join}, the three-class, two-queue system is stable if and only if
$$
\max\{a(1-d)(1-p)\lambda/\mu,
(1-a(1-d))\lambda/\mu,
\lambda/(2\mu)\}<1,
$$
which is equivalent to \eqref{eq_stable1}--\eqref{eq_stable3}.
\hfill$\square$

\begin{figure}[htb]
\centering
\includegraphics[width=0.4\textwidth]{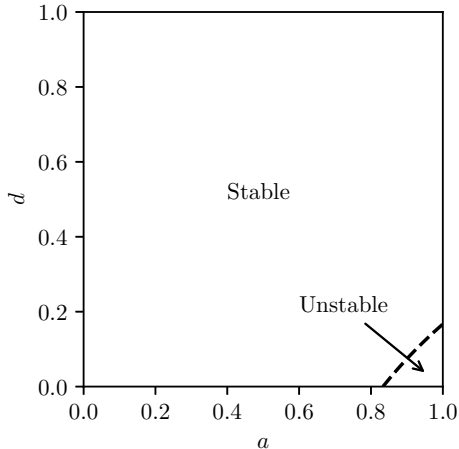}
\caption{Stability of shorter-queue system for various attacking probability $a$ and defending probability $d$.}
\label{fig_stabilityShorterQueue}
\end{figure}

Proposition \ref{prp_1} indicates the two-queue system with shorter-queue routing is stable regardless of attack and defence given $\lambda < \mu$. But when $\lambda \ge \mu$, the system stability is associated with $a$, $d$ and $p$. For example, Fig.~\ref{fig_stabilityShorterQueue} illustrates the stability regime in the $a$-$d$ plane given $\lambda=0.6$, $\mu=0.5$, $p=1$. It illustrates defending probability $d$ should not be too low when attacking probability is high; otherwise, the system is unstable.

\subsection{Computing queuing cost}

Analytical solution to the shorter-queue problem is very hard and unnecessarily complex for our purpose. Instead, we derive a simple theoretical upper bound as an approximation. The derivation is based on \cite{foley2001join}.

\begin{prp}
The mean number of jobs $\bar N=\bar X+\bar Y$ in the system is upper bounded by
\begin{align}
    \bar N &\le \bar n(a,p,d;\lambda) \nonumber \\
    &:=-2+\frac{2\mu}{\min\{\mu-\tilde{a}p\lambda,\mu-\tilde{a}(1-p)\lambda,\mu-\lambda/2\}}
    \label{Nbar}
\end{align}
where $\tilde{a} = a(1-d)$.
\end{prp}
\section{Security game}
\label{sec_game}

Consider the two-queue system with the shorter-queue routing policy and under the attack of the attacker and defense of the operator (defender).
Throughout this paper, we assume that the system is nominally stable, i.e.
$$
\lambda<2\mu.
$$

An attack induces a technological cost of $c_a\ge0$ on the attacker.
The attacker's utility is the difference of the (upper bound of) average total queue size and the average technological cost:
$$
u_a(a,p,d;\lambda)=\bar n(a,p,d;\lambda) - \lambda c_aa.
$$

As for the operator, protecting a job induces a technological cost of $c_d \ge 0$. The operator aims at shorter queue length and lower cost, and thus the utility is given by
$$
u_d(a,p,d;\lambda)=-\bar n(a,p,d;\lambda) - \lambda c_dd.
$$

An important observation from \eqref{Nbar} is that the best response of the attacker must be such that either $p=0$ or $p=1$. That is, when the attacker modifies the routing instruction, the attacker always allocate jobs to the same server. This is intuitive in that sending all jobs to one server will cause higher delay than distributing jobs over two servers. Hence, $p$ can be actually dropped from the utility function, and we will let $p=1$ henceforth. With a slight abuse of notation, we write
\begin{align*}
     u_a(a,d;\lambda)
     =\begin{cases}
     \!-\!2\!+\!\frac{2\mu}{\min\{\mu-\tilde{a}\lambda,\mu-\lambda/2\}}-\lambda c_aa & \tilde{a}<\frac{\mu}{\lambda}\\
     +\infty & \tilde{a}\ge \frac{\mu}{\lambda}
     \end{cases}
\end{align*}
and
 \begin{align*}
 u_d(a,d;\lambda)&=\begin{cases}
     2-\frac{2\mu}{\min\{\mu-\tilde{a}\lambda,\mu-\lambda/2\}}-\lambda c_dd & \tilde{a}<\frac{\mu}{\lambda}\\
     -\infty & \tilde{a}\ge \frac{\mu}{\lambda}
     \end{cases}
 \end{align*}
where $\tilde{a}=a(1-d)$.

\begin{figure}[ht]
\centering
\begin{subfigure}{0.32\textwidth}
  \centering
  \includegraphics[width=\linewidth]{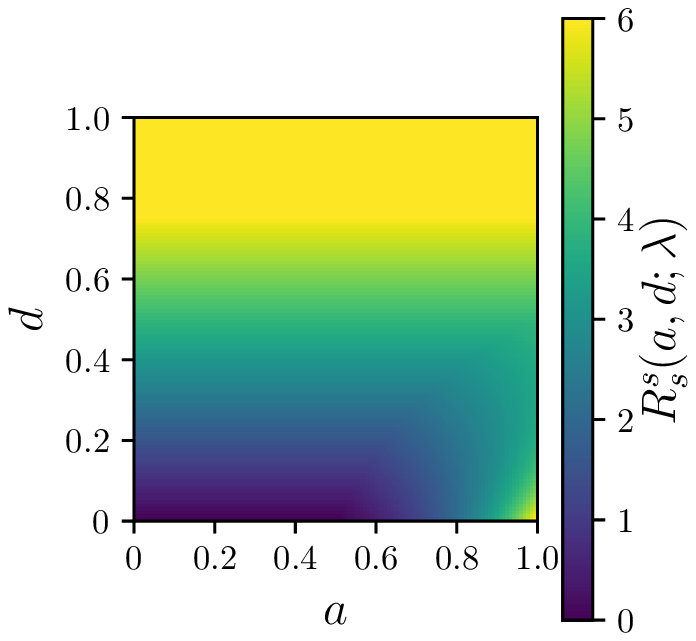}  
  \caption{$\lambda=0.4$}
  \label{fig_Rs_0.4}
\end{subfigure}
\begin{subfigure}{0.32\textwidth}
  \centering
  \includegraphics[width=\linewidth]{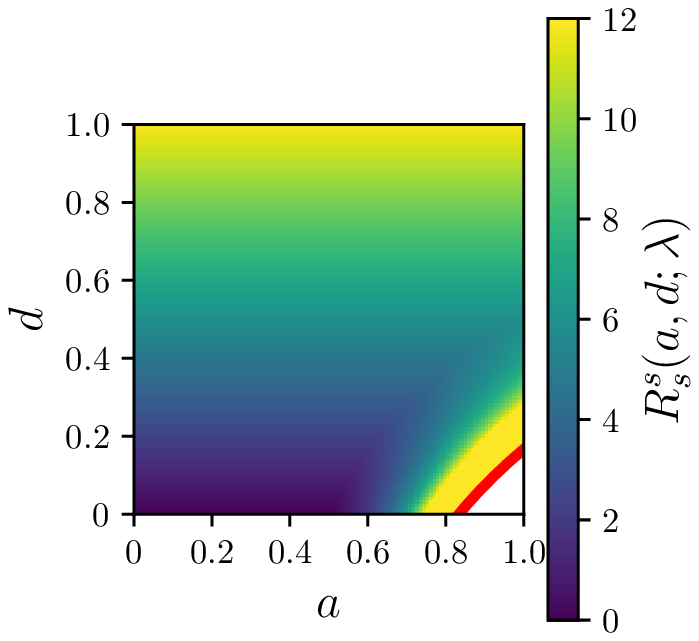}  
  \caption{$\lambda=0.6$}
  \label{fig_Rs_0.6}
\end{subfigure}
\caption{Security risk under shorter-queue routing}
\label{fig_Rs}
\end{figure}

We define security risk $R_s^s(a,d;\lambda)$ for two-queue system with shorter-queue routing as follows
$$R_s^s(a,d;\lambda):=u_d(0, 0;\lambda)-u_d(a, d;\lambda).$$
Fig.~\ref{fig_Rs} illustrates $R_s^s(a,d;\lambda)$ in two numerical examples with different $\lambda$, given $\mu=0.5$ and $c_d=20$. Fig.~\ref{fig_Rs_0.4} shows that $R_s^s(a, d; \lambda)$ mainly rises with $d$, which means that defense cost $c_d$ dominates in security risk when $\lambda=0.4$ and $\mu=0.6$. In this case, it is expected that defense might be forsaken if defense cost becomes too high. In addition, Fig.~\ref{fig_Rs_0.6} shows the stronger relationship between $R_s^s(a,d;\lambda)$ and attack probability $a$ when $\lambda=0.6$ and $\mu=0.5$. Under the fierce attack, $R_s^s(a,d;\lambda)$ increases dramatically, and the operator must take defense action to minimize the security risk; otherwise, the system would get into the unstable state that is denoted by the empty under red line in Fig.~\ref{fig_Rs_0.6}.

We use $(a^*, d^*)$ to denote the equilibria in the above security game, then we have the following theorem. 
 \begin{thm}
 \label{thm}
The attacker-defender game has the following regimes of equilibria:
\begin{enumerate}
    \item[(A)]Regime A: $a^*=0,\ d^*=0$;
    \item[(B)]Regime B: $a^*=1$. According to the value of $d^*$, this regime can be further partitioned into two subregimes:
\begin{enumerate}
    \item[($B_1$)] $a^*=1,\ d^*=0$;
    \item[($B_2$)] $a^*=1,\ d^*=1-\frac1\lambda(\mu-\sqrt{2\mu/c_d})$.
\end{enumerate}
\end{enumerate}
Furthermore, regime $B_1$ is non-empty if and only if $\lambda<\mu$.
\end{thm}

The rest of this section is devoted to the proof of Theorem~\ref{thm} and the characterization and visualization of the regime boundaries.

\subsection{Properties of equilibria}

Any equilibrium $(a^*,d^*)$ satisfies
\begin{align*}
    &a^*=\arg\max_{a\in[0,1]} \bar n(a,d^*;\lambda)-\lambda c_aa,\\
    &d^*=\arg\max_{d\in[0,1]} -\bar n(a^*,d;\lambda)-\lambda c_dd.
\end{align*}

The following results characterize important qualitative properties of the equilibria, which are the basis for our subsequent analysis.

\begin{prp}\label{a=0or1}
For any equilibrium $(a^*,d^*)$, either $a^*=0$ or $a^*=1$.
\end{prp}
\noindent\emph{Proof.}
Given $d=d^*$, the attacker's utility is given by
\begin{align*}
    u_a(a,d^*;\lambda)=-2+\frac{2\mu}{\min\{\mu-a(1-d^*)\lambda,\mu-\lambda/2\}}-\lambda c_aa.
\end{align*}
We need to consider two cases. In the first case that $\mu-a(1-d^*)\lambda >\mu-\lambda/2$, we have
\begin{align*}
    u_a(a,d^*;\lambda)=-2+\frac{2\mu}{\mu-\lambda/2}-\lambda c_aa,
\end{align*}
which immediately imply that $a^*=\arg\max_{a\in[0,1]}f(a)=0$.
In the second case that $\mu-a(1-d^*)\lambda <\mu-\lambda/2$, note that the stability condition
$\mu-a(1-d^*\lambda)>0$ must hold; otherwise the defender must be able to improve the utility by increasing $d$. Thus, we have
\begin{align*}
    &u_a(a,d^*;\lambda)=-2+\frac{2\mu}{\mu-a(1-d^*)\lambda}-\lambda c_aa,\\
    &\frac{\partial^2}{\partial a^2} u_a(a,d^*;\lambda)=\frac{2\lambda(1-d^*)}{[\mu-a(1-d^*)\lambda]^3}
\end{align*}
which implies that $u_a(a,d^*;\lambda)$ is convex in $a$ for any $d^*$; therefore $a^*=0$ or $a^*=1$.
\hfill$\square$

\begin{prp}\label{a=0}
For any equilibrium $(a^*,d^*)$ such that $a^*=0$, we have $d^*=0$.
\end{prp}
\noindent\emph{Proof.}
If $a^*=0$, we have
\begin{align*}
d^*=\arg\max_{d\in[0,1]}-2+\frac{2\mu}{\mu-\lambda/2}+\lambda c_dd,
\end{align*}
which immediately implies that $d^*=0$.
\hfill$\square$

\subsection{Regime boundaries}
\label{sub_regimes}

Since, by Proposition~\ref{a=0or1}, each equilibrium $(a^*,d^*)$ satisfies either $a^*=0$ or $a^*=1$, we only need to consider the utilities for $a=0$ and $a=1$.
By Proposition~\ref{a=0}, the best response for the defender when $a=0$ is $d^*(0)=0$. For $a=0$ and $d=0$, we have
\begin{align*}
    &u_a(0,0;\lambda)=-2+\frac{2\mu}{\mu-\lambda/2},\\
    &u_d(0,0;\lambda)=2-\frac{2\mu}{\mu-\lambda/2}.
\end{align*}

For $a=1$, we have
\begin{align*}
    &u_a(1,d;\lambda)\!=\!\begin{cases}
    \!-\!2\!+\!\frac{2\mu}{\min\{\mu-(1-d)\lambda,\mu-\lambda/2\}}\!-\!\lambda c_a & 1-d<\frac{\mu}{\lambda}\\
    +\infty & 1-d\ge\frac{\mu}{\lambda},
    \end{cases}\\
    &u_d(1,d;\lambda)\!=\!\begin{cases}
    2\!-\!\frac{2\mu}{\min\{\mu-(1-d)\lambda,\mu-\lambda/2\}}\!-\!\lambda c_dd & 1-d<\frac{\mu}{\lambda}\\
    -\infty & 1-d\ge\frac{\mu}{\lambda}.
    \end{cases}
\end{align*}

For ease of presentation, define
\begin{align}
    \gamma:=\frac1\lambda\sqrt{\mu}(\sqrt{\mu}-\sqrt{2/c_d}).
\end{align}

Given $a=1$, the best response for the defender is given by
\begin{align*}
    d^*(1)=\begin{cases}
    0 & \mbox{if $\gamma\ge1$}\\
    1-\gamma
    &\mbox{if $1/2<\gamma<1$}\\
    \frac12 & \mbox{if $\gamma\le1/2$.}
    \end{cases}
\end{align*}
and the utility associated with the above best response is given by
\begin{align*}
    u_a(1,d^*(1);\lambda)=\begin{cases}
    -2+\frac{2\mu}{\mu-\lambda}-\lambda c_a & \mbox{if $\gamma\ge1$}\\
    -2+\frac{2\mu}{\mu-\gamma\lambda}-\lambda c_a
    &\mbox{if $1/2<\gamma<1$}\\
    -2+\frac{2\mu}{\mu-\lambda/2}-\lambda c_a & \mbox{if $\gamma\le1/2$.}
    \end{cases}
\end{align*}

For $\gamma\le1/2$, we have $u_a(0,d^*(0);\lambda)>u_a(1,d^*(1);\lambda)$. Therefore, the equilibrium is $(0,0)$.

For $1/2<\gamma<1$, if
$$
-2+\frac{2\mu}{\mu-\lambda/2}>
-2+\frac{2\mu}{\mu-\gamma\lambda}-\lambda c_a,
$$
the equilibrium is $(0,0)$,
and if
$$
-2+\frac{2\mu}{\mu-\lambda/2}<
-2+\frac{2\mu}{\mu-\gamma\lambda}-\lambda c_a,
$$
the equilibrium is $(1,1-\gamma)$.

For $\gamma\ge1$, if
$$
-2+\frac{2\mu}{\mu-\lambda/2}>
-2+\frac{2\mu}{\mu-\lambda}-\lambda c_a,
$$
the equilibrium is $(0,0)$,
and if
$$
-2+\frac{2\mu}{\mu-\lambda/2}<
-2+\frac{2\mu}{\mu-\lambda}-\lambda c_a,
$$
the equilibrium is $(1,0)$.

In summary, the regimes are
\begin{enumerate}
    \item[($A$)] $a^*=0,\ d^*=0$ if
    (i) $1/2<\gamma<1$ and $\frac{2\mu}{\mu-\lambda/2}>
\frac{2\mu}{\mu-\gamma\lambda}-\lambda c_a
$, or if 
    (ii) $\gamma\ge1$ and $\frac{2\mu}{\mu-\lambda/2}>
\frac{2\mu}{\mu-\lambda}-\lambda c_a
$, or if 
(iii) $\gamma<1/2$;
    \item[($B$)] $a^*=1$ with two subregimes:
\begin{enumerate}
    \item[($B_1$)] $a^*=1,\ d^*=0$ if $
\frac{2\mu}{\mu-\lambda/2}<
\frac{2\mu}{\mu-\lambda}-\lambda c_a
$ and $\gamma\ge1$;
    \item[($B_2$)] $a^*=1,\ d^*=1-\frac1\lambda(\mu-\sqrt{2\mu/c_d})$ if $
\frac{2\mu}{\mu-\lambda/2}<
\frac{2\mu}{\mu-\gamma\lambda}-\lambda c_a
$ and $1/2<\gamma<1$.
\end{enumerate}
\end{enumerate}

\begin{figure}[ht]
\centering
\begin{subfigure}{0.32\textwidth}
  \centering
  \includegraphics[width=\linewidth]{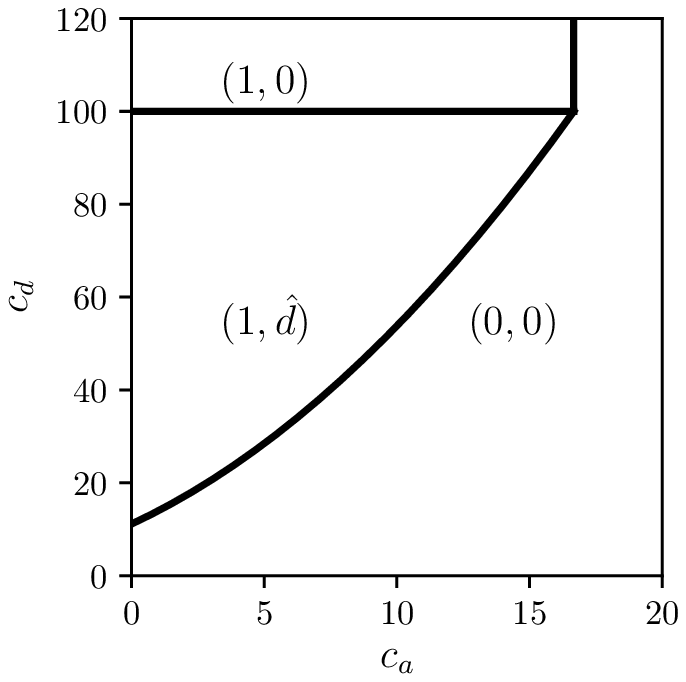}  
  \caption{$\lambda=0.4$}
  \label{fig_regime_shorter_0.4}
\end{subfigure}
\begin{subfigure}{0.32\textwidth}
  \centering
  \includegraphics[width=\linewidth]{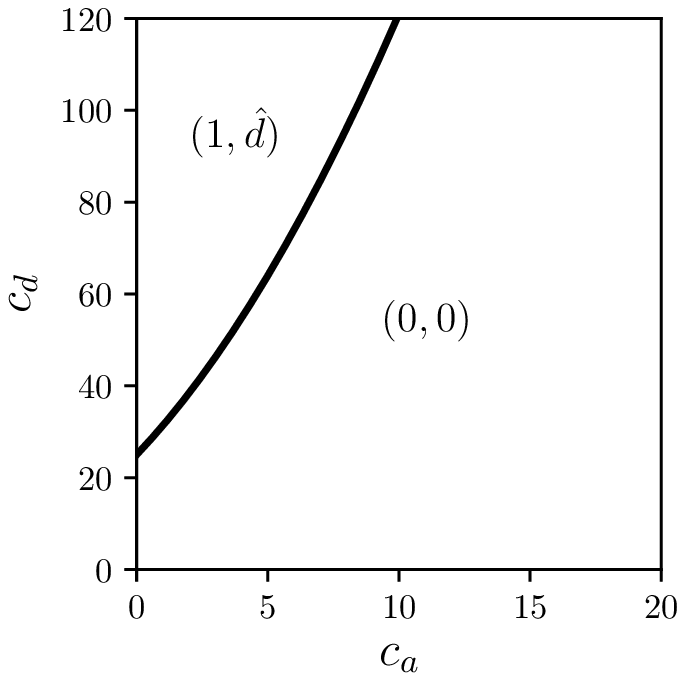}  
  \caption{$\lambda=0.6$}
  \label{fig_regime_shorter_0.6}
\end{subfigure}
\caption{Equilibrium regime under Shorter routing}
\label{fig_regime_shorter}
\end{figure}

Fig.~\ref{fig_regime_shorter} illustrates the regimes in two numerical cases with different $\lambda$, given $\mu=0.5$. Each regime is labeled with the corresponding $(a^*, d^*)$. $(0, 0)$ indicates that there is no attack and therefore no defense is needed. This regime is associated with high attack cost $c_a$ and low defense cost $c_d$. Given large $c_a$, the attack has no incentive and given small $c_d$, the attack must be counteracted by the defender. When $c_d$ increases, the defender's strategy will be increasingly influenced by the technological cost, which leaves opportunities for the attacker. This regime is denoted by $(1, \hat{d})$. Importantly, the defender's action strongly depend on whether $\lambda$ is less than $\mu$. As shown in Fig.~\ref{fig_regime_shorter_0.4}, the attacker has no incentive to defend when $\lambda < \mu$. But given $\lambda \ge \mu$, the defense would continue since the outcome of instability is much severer. As a result, $(1, 0)$ is removed in Fig.~\ref{fig_regime_shorter_0.6}.

\section{Comparison with open-loop routing}
\label{sec_compare}

\begin{figure}[htb]
\centering
\includegraphics[width=8cm]{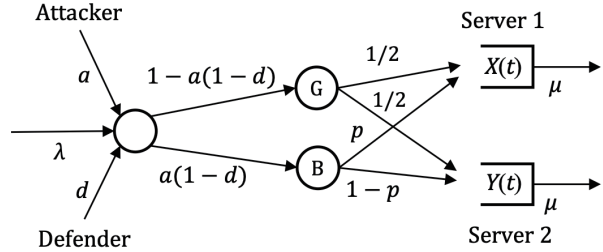}
\caption{Two-queue system with Bernoulli routing and malicious attacks.}
\label{fig_twoqueue2}
\end{figure}

To evaluate the system performance under shorter-queue routing, we compare it with that under Bernoulli routing. Bernoulli routing herein means that the router assigns jobs to each server with probability 1/2; see Fig.~\ref{fig_twoqueue2}. We first point out the queuing cost for Bernoulli routing, then compare the security risk and equilibrium regime with those of shorter-queue routing.

\subsection{Queuing cost for Bernoulli routing}
The job arriving at server 1 can be divided into two classes. The first is attacked, while the second is not attacked. The arrival rates are $pa(1-d)\lambda$ and $(1-a(1-d))\lambda/2$ respectively. Thus the arrival rate at server 1 equals $(\frac{1-a(1-d)}{2}+pa(1-d))\lambda$. Recall that we use $\tilde{a}$ to denote $a(1-d)$. Then the arrival rate at server 1 is simplified as $\frac{1-\tilde{a}+2p\tilde{a}}{2}\lambda$, and, by standard results in queuing theory \cite[p.326]{gallager2013stochastic} we have the mean number of jobs at server 1

$$\bar{X} = \frac{(1-\tilde{a}+2p\tilde{a})\lambda}{2\mu-(1-\tilde{a}+2p\tilde{a})\lambda}.$$

The mean number of jobs at server 2 is computed similarly, and we have the total job number:

$$\bar{X}+\bar{Y} =\frac{(1-\tilde{a}+2p\tilde{a})\lambda}{2\mu-(1-\tilde{a}+2p\tilde{a})\lambda}+\frac{(1+\tilde{a}-2p\tilde{a})\lambda}{2\mu-(1+\tilde{a}-2p\tilde{a})\lambda}.
$$

\subsection{Security game for Bernoulli routing}
Similar to the attack on shorter-queue routing, the best attack strategy must be either $p=1$ or $p=0$. We let $p=1$, then the attacker's utility under Bernoulli routing is
\begin{align*}
     v_a(a,d;\lambda)
     =\begin{cases}
     \frac{(1+\tilde{a})\lambda}{2\mu-(1+\tilde{a})\lambda}
+\frac{(1-\tilde{a})\lambda}{2\mu-(1-\tilde{a})\lambda}
-c_a\lambda a & 1+\tilde{a}<\frac{2\mu}{\lambda}\\
     +\infty & 1+\tilde{a}\ge \frac{2\mu}{\lambda},
     \end{cases}
\end{align*}
and the defender's utility is
\begin{align*}
v_d(a,d;\lambda)
     =\begin{cases}
     \frac{-(1+\tilde{a})\lambda}{2\mu-(1+\tilde{a})\lambda}
-\frac{(1-\tilde{a})\lambda}{2\mu-(1-\tilde{a})\lambda}
-c_d\lambda d & 1+\tilde{a}<\frac{2\mu}{\lambda}\\
     -\infty & 1+\tilde{a}\ge \frac{2\mu}{\lambda}.
     \end{cases}
\end{align*}

We define security risk $R_s^b(a,d;\lambda)$ for two-queue system with Bernoulli routing as follows
$$R_s^b(a,d;\lambda):=v_d(0, 0;\lambda)-v_d(a, d;\lambda).$$

\begin{figure}[ht]
\centering
\begin{subfigure}{0.32\textwidth}
  \centering
  \includegraphics[width=\linewidth]{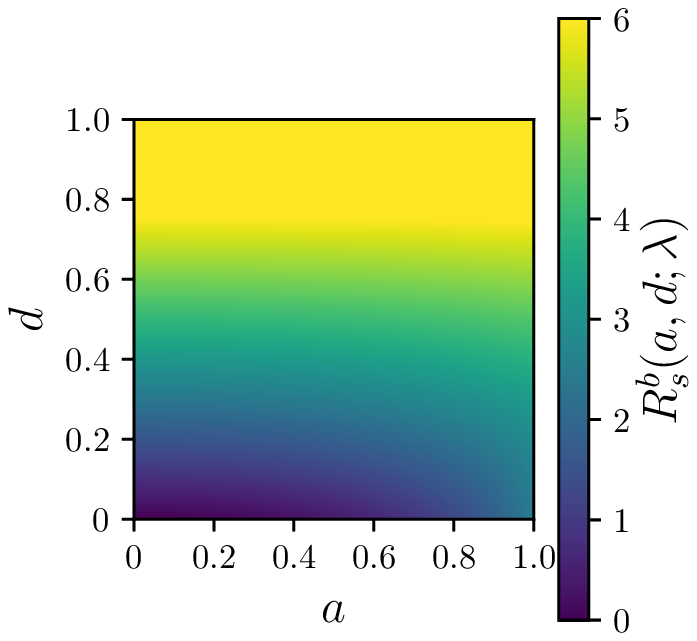}  
  \caption{$\lambda=0.4$}
  \label{fig_Rb_0.4}
\end{subfigure}
\begin{subfigure}{0.32\textwidth}
  \centering
  \includegraphics[width=\linewidth]{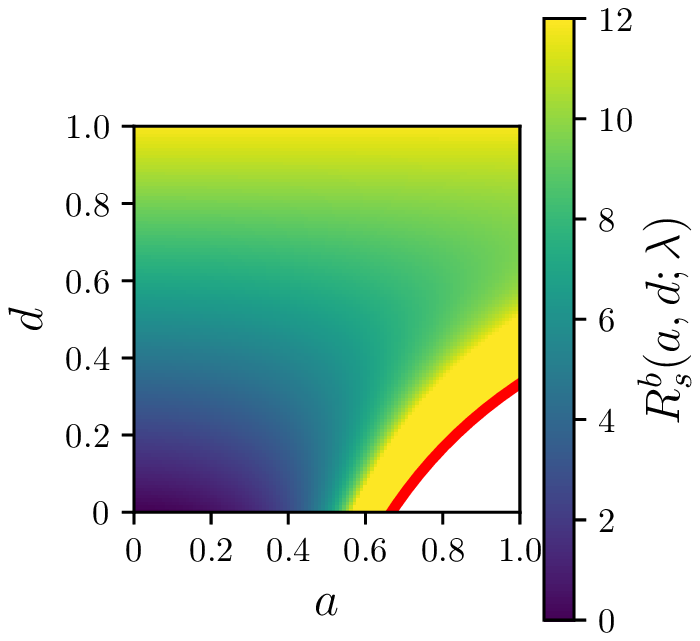}  
  \caption{$\lambda=0.6$}
  \label{fig_Rb_0.6}
\end{subfigure}
\caption{Security risk under Bernoulli routing}
\label{fig_Rb}
\end{figure}

Fig.~\ref{fig_Rb} illustrates $R_s^b(a,d;\lambda)$ in two numerical examples with different $\lambda$, given $\mu=0.5$ and $c_d=20$. The revealed relationship between security risk and $a$, $d$ in Bernoulli routing is similar to that in shorter-queue routing. However, in terms of security risk, Fig.~\ref{fig_Rs_0.4} and Fig.~\ref{fig_Rb_0.4} show Bernoulli routing slightly superior to shorter-queue routing, while Fig.~\ref{fig_Rs_0.6} and Fig.~\ref{fig_Rb_0.6} show shorter-queue routing is much better than Bernoulli routing. Combining Fig.~\ref{fig_Rs_0.4} and Fig.~\ref{fig_Rb_0.4}, we find $R_s^b(a, d; \lambda)$ is lower than $R_s^s(a, d; \lambda)$ when $(a, d)$ is around $(1, 0)$ given $\lambda=0.4$. On the contrary, Fig.~\ref{fig_Rs_0.6} and  Fig.~\ref{fig_Rb_0.6} show security risks of Bernoulli routing is obviously higher that those of shorter-queue routing when $a>0.5$, given $\lambda=0.6$. 

The contradictory observations from Fig.~\ref{fig_Rs} and Fig.~\ref{fig_Rb} is caused by the approximation error of queue cost in shorter-queue routing. Recall that we use a up bound to represent the queue cost for shorter-queue routing, and the above findings indicate the upper bound might overestimate the queue length when $\lambda < \mu$ and then mislead the strategy in equilibrium, which is demonstrated later.

We use $(a^\dagger, d^\dagger)$ to denote the equilibrium in the security game for Bernoulli routing. The regimes are summarized as follows and more details are available in the appendix.
\begin{enumerate}
    \item[($A$)] $a^\dagger=0,\ d^\dagger=0$ if (i) $\lambda < \mu$, $\frac{(2\mu-\lambda)\lambda}{2(\mu-\lambda)\mu} \le c_d$ and $\frac{2\lambda}{2\mu-\lambda} \geq \frac{\lambda}{\mu-\lambda} - c_a\lambda$, or if (ii) $\lambda < \mu$, $\frac{(2\mu-\lambda)\lambda}{2(\mu-\lambda)\mu} > c_d$ and $\frac{2\lambda}{2\mu-\lambda}\geq \frac{(2-\hat{d})}{2\mu-(2-\hat{d})\lambda} + \frac{\hat{d}\lambda}{2\mu-\hat{d}\lambda}-c_a \lambda$, or if (iii) $\lambda \ge \mu$ and $\frac{2\lambda}{2\mu-\lambda} \geq \frac{(2-\hat{d})\lambda}{2\mu-(2-\hat{d})\lambda} + \frac{\hat{d}\lambda}{2\mu-\hat{d}\lambda} - c_a \lambda$;
    \item[($B$)] $a^\dagger=1$ with two subregimes:
    \begin{enumerate}
        \item[($B_1$)] $a^\dagger=1$, $d^\dagger=0$ if $\lambda < \mu$, $\frac{(2\mu-\lambda)\lambda}{2(\mu-\lambda)\mu} \le c_d$ and $\frac{2\lambda}{2\mu-\lambda} < \frac{\lambda}{\mu-\lambda}-c_a\lambda$
        \item[($B_2$)]
        $a^\dagger=1$, $d^\dagger=\hat{d}$ if (i) $\lambda < \mu$, $\frac{(2\mu-\lambda)\lambda}{2(\mu-\lambda)\mu} > c_d$ and $\frac{2\lambda}{2\mu-\lambda} < \frac{(2-\hat{d})}{2\mu-(2-\hat{d})\lambda} + \frac{\hat{d}\lambda}{2\mu-\hat{d}\lambda}-c_a \lambda$, or if (ii) $\lambda \ge \mu$ and $\frac{2\lambda}{2\mu-\lambda} < \frac{(2-\hat{d})\lambda}{2\mu-(2-\hat{d})\lambda} + \frac{\hat{d}\lambda}{2\mu-\hat{d}\lambda} - c_a \lambda$.
    \end{enumerate}
\end{enumerate}
where $\hat{d} = 1 - \frac{1}{\lambda}(\theta-\sqrt{\zeta^2-\theta^2+\frac{2\kappa\zeta}{\theta}})$, $\zeta=2\mu-\lambda$, $\kappa=\mu/c_d$,  $\theta=\sqrt{\eta+\frac{\zeta^4}{9\eta}+\frac{\zeta^2}{3}}$, $\eta = \sqrt[3]{\frac{\zeta^6}{27} + \frac{\kappa^2\zeta^2}{2} + \sqrt{\frac{\kappa^2\zeta^8}{27}+\frac{\kappa^4\zeta^4}{4}}}$.

\begin{figure}[ht]
\centering
\begin{subfigure}{0.32\textwidth}
  \centering
  \includegraphics[width=\linewidth]{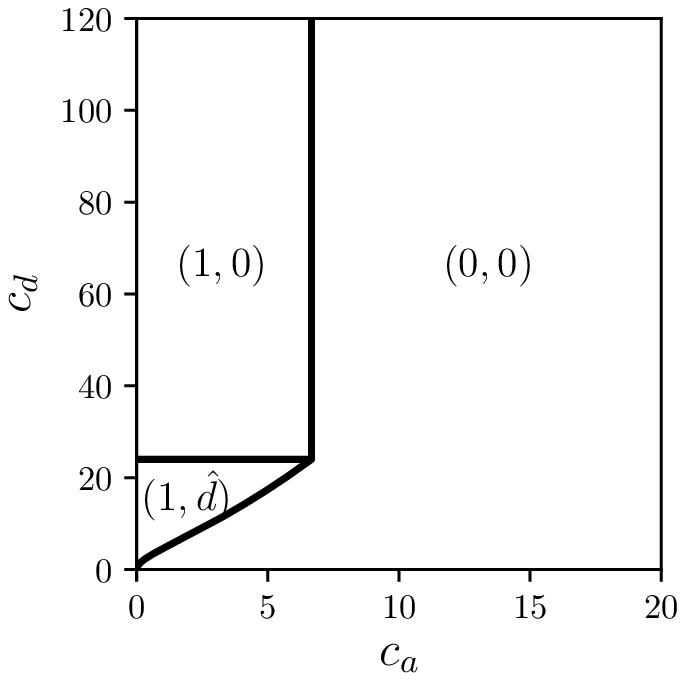}  
  \caption{$\lambda=0.4$}
  \label{fig_regime_bernoulli_0.4}
\end{subfigure}
\begin{subfigure}{0.32\textwidth}
  \centering
  \includegraphics[width=\linewidth]{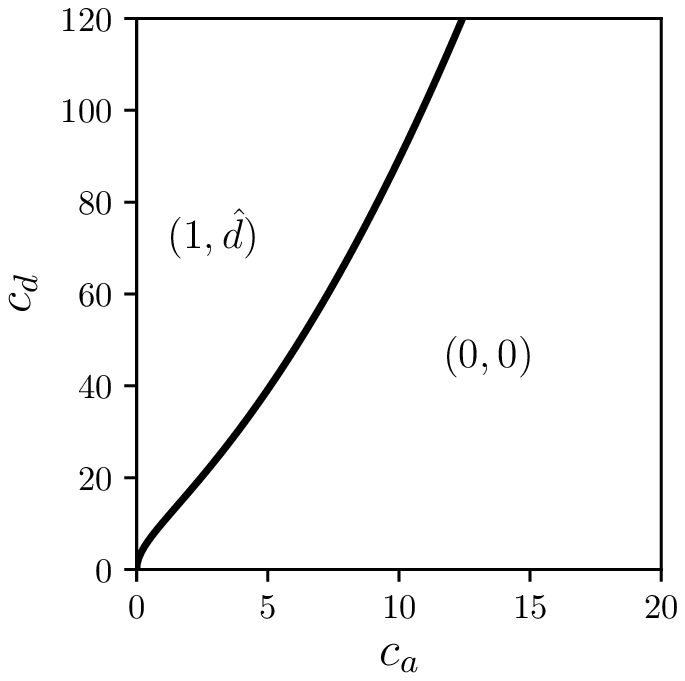}  
  \caption{$\lambda=0.6$}
  \label{fig_regime_bernoulli_0.6}
\end{subfigure}
\caption{Equilibrium regime under Bernoulli routing}
\label{fig_regime_bernoulli}
\end{figure}

Fig.~\ref{fig_regime_bernoulli} illustrates the regimes in two numerical examples with different $\lambda$, given $\mu=0.5$. Each regime is labeled with the corresponding $(a^\dagger, d^\dagger)$. Compared with Fig. \ref{fig_regime_shorter_0.4}, Fig.~\ref{fig_regime_bernoulli_0.4} shows that $(1, 0)$ can be achieved with lower $c_d$ given $c_a$ and $(0, 0)$ can be also realized with lower $c_a$ given $c_d$. The reason is that the queue cost of shorter-queue routing is overestimated by the upper bound and thus the attacker/defender is willing to undertake more attack/defense costs. In case of $\lambda=0.6$, Fig.~\ref{fig_regime_bernoulli_0.6} illustrates that $(0, 0)$ is achieved with larger $c_a$ given $c_d$. This means Bernoulli routing is less effective than shorter-queue routing since it requires larger $c_a$ to prevent attack.

\begin{figure}[ht]
\centering
\begin{subfigure}{0.32\textwidth}
  \includegraphics[width=\linewidth]{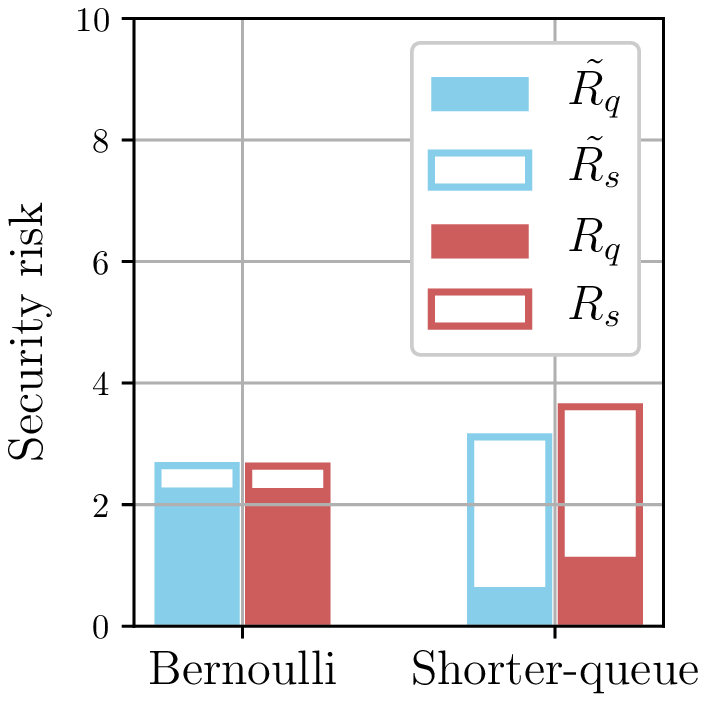}  
  \caption{$\lambda=0.4$, $c_d=20$}
  \label{fig_comp_1}
\end{subfigure}
\begin{subfigure}{0.32\textwidth}
  \includegraphics[width=\linewidth]{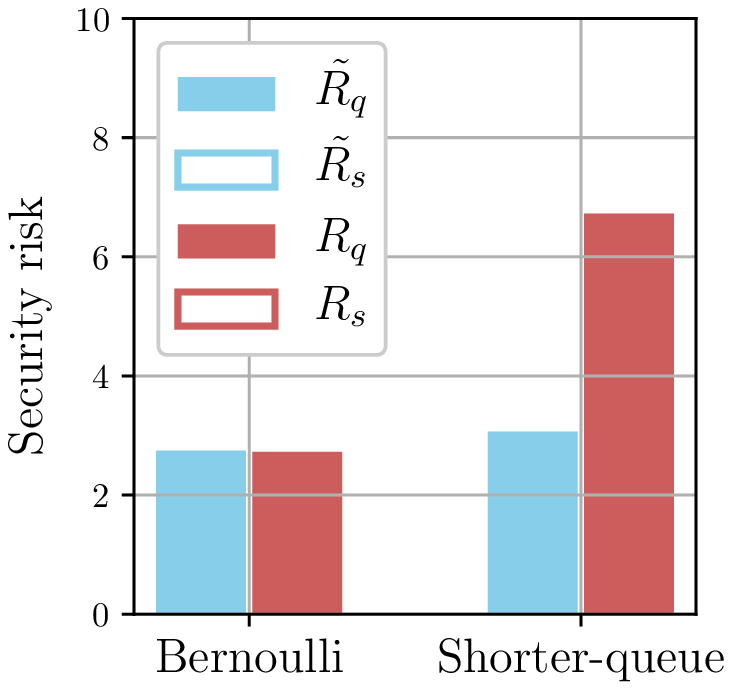}  
  \caption{$\lambda=0.4$, $c_d=110$}
  \label{fig_comp_2}
\end{subfigure}
\begin{subfigure}{0.32\textwidth}
  \includegraphics[width=\linewidth]{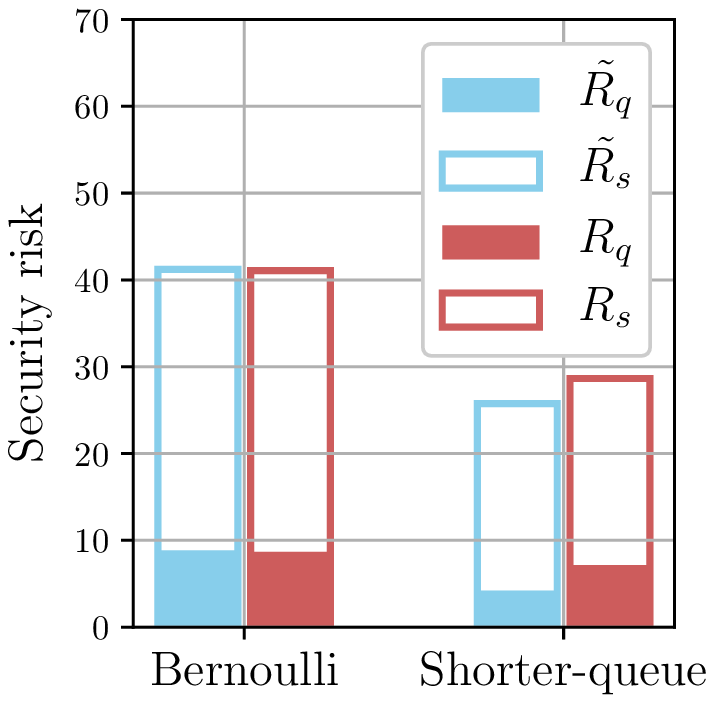}  
  \caption{$\lambda=0.6$, $c_d=110$}
  \label{fig_comp_3}
\end{subfigure}
\caption{Comparison of security risk}
\label{fig_comp}
\end{figure}

We further compare security risks in a state of equilibrium under the two routing policies. By fixing $\mu=0.5$ and $c_a=1$, we choose $(\lambda, c_d)=(0.4, 20), (0.4, 110), (0.6, 110)$ to study three kinds of equilibrium respectively, namely $(1, \hat{d})$ given $\lambda < \mu$, $(1, 0)$ given $\lambda < \mu$ and $(1, \hat{d})$ given$\lambda \ge \mu$. The comparison is presented in Fig.~\ref{fig_comp}, where $R_q$ and $R_s$ respectively denote queue risk and security risk through theoretical analysis, while $\tilde{R_q}$ and $\tilde{R_s}$ denote the values through numerical simulation. Herein we define queue risk $R_q$ as queue length under attack and defense minus that free from attack. Then the blank area in the bars of Fig.~\ref{fig_comp} can be recognized as defense costs. Fig.~\ref{fig_comp_1} shows the defense cost of shorter-queue routing in equilibrium given $\lambda=0.4, \mu=0.5, c_a=1, c_d=20$ is much more than that of Bernoulli routing. The reason might lie in that the queue length overestimated by the upper bound induces the defender to adopt larger $\hat{d}$, which finally results in more security risk. Fig.~\ref{fig_comp_2} explicitly presents the relative approximation error might be large when $\lambda < \mu$. Fig.~\ref{fig_comp_3} illustrates the huge advantage of shorter-queue routing when $\lambda \ge \mu$. The numerical simulation shows that the security risk is decreased by 37\%.

\section{Concluding remarks}
This work quantifies the security risks of two-queue system that is routed by shortest-queue policy and suffers malicious attack. Our theoretical analysis can help decision-makers figure out appropriate strategies against attack. The comparison with Bernoulli routing demonstrates that the proposed methodology has great potentials, especially when the system is congested, but it also indicates the requirement for more powerful tools that accurately approximates queue cost in two-queue system. 

This work can serve as the basis for multiple future research directions. First, the impact of more sophisticated attacking strategies such as state-dependent attacking probability can be studied in our framework. Second, the effect of fault-tolerant routing algorithms can be analyzed in terms of reducing security risks. Third, multi-stage attacker-defender game with strategy learning (e.g. the formulation in \cite{wu2019learning}) may provide additional insights about secure design.
\section*{Appendix}

\subsection{Equilibrium regime under Bernoulli routing}
Like equilibria $(a^*, d^*)$ in shorter-queue routing, equilibria $(a^\dagger, d^\dagger)$ have the properties that $a^\dagger$ equals either 0 or 1 and $d^\dagger$ must equal 0 given $a^\dagger=0$. For $a=0$ and $d=0$, we have
\begin{align*}
    v_a(0, 0; \lambda) &= \frac{2\lambda}{2\mu-\lambda},\\
    v_d(0, 0; \lambda) &= -\frac{2\lambda}{2\mu-\lambda}.
\end{align*}

For $a=1$, the attacker's utility is given by
\begin{align*}
     v_a(1,d;\lambda)
     &=\begin{cases}
     \frac{(2-d)\lambda}{2\mu-(2-d)\lambda}
+\frac{d\lambda}{2\mu-d\lambda}
-c_a\lambda & 2-d<\frac{2\mu}{\lambda}\\
     +\infty &  2-d\ge \frac{2\mu}{\lambda},
     \end{cases}
\end{align*}
and the defender's utility is given by
\begin{align*}
v_d(1,d;\lambda)
     &=\begin{cases}
     \frac{-(2-d)\lambda}{2\mu-(2-d)\lambda}
-\frac{d\lambda}{2\mu-d\lambda}
-c_d\lambda d & 2-d<\frac{2\mu}{\lambda}\\
     -\infty & 2-d\ge \frac{2\mu}{\lambda}.
     \end{cases}
\end{align*}

Given $a=1$, we have the following best response
\begin{align*}
    d^*(1)=\begin{cases}
    0 & \mbox{if $\lambda<\mu$ and $\frac{(2\mu-\lambda)\lambda}{2(\mu-\lambda)\mu} - c_d > 0$}\\
    \hat{d}
    &\mbox{if i) $\lambda<\mu$ and $\frac{(2\mu-\lambda)\lambda}{2(\mu-\lambda)\mu} - c_d \le 0$  or ii) $\lambda \ge \mu$}
    \end{cases}
\end{align*}
where $\frac{(2\mu-\lambda)\lambda}{2(\mu-\lambda)\mu} - c_d$ denotes $\frac{\partial v_d(1,d;\lambda)}{\partial d}\big|_{d = 0}$ and $\hat{d}$ satisfies $\frac{\partial v_d(1,d;\lambda)}{\partial d}\big|_{d=\hat{d}} = 0$.

By solving $\frac{\partial v_d(1,d;\lambda)}{\partial d} = 0$, we have
$$\hat{d} = 1 - \frac{1}{\lambda}(\theta-\sqrt{\zeta^2-\theta^2+\frac{2\kappa\zeta}{\theta}})
$$
where $\zeta=2\mu-\lambda$, $\kappa=\mu/c_d$,  $\theta=\sqrt{\eta+\frac{\zeta^4}{9\eta}+\frac{\zeta^2}{3}}$, $\eta = \sqrt[3]{\frac{\zeta^6}{27} + \frac{\kappa^2\zeta^2}{2} + \sqrt{\frac{\kappa^2\zeta^8}{27}+\frac{\kappa^4\zeta^4}{4}}}$.

The attacker's best utility is given by
\begin{align*}
& u_a(1,d^*(1);\lambda)\\
&= \left\{
\begin{array}{ll}
     \frac{\lambda}{\mu-\lambda}-\lambda c_a & \mbox{if $\lambda<\mu$ and $\frac{(2\mu-\lambda)\lambda}{2(\mu-\lambda)\mu} > c_d$}  \\
     \frac{(2-\hat{d})\lambda}{\mu-(2-\hat{d})\lambda} + \frac{\hat{d}\lambda}{\mu-\hat{d}\lambda} -\lambda c_a  & \mbox{if i)$\lambda<\mu$ and $\frac{(2\mu-\lambda)\lambda}{2(\mu-\lambda)\mu} \le c_d$} \\
     & \mbox{or ii)$\lambda \ge\mu$}.
\end{array}
\right.
\end{align*}

For $\lambda < \mu$ and $\frac{(2\mu-\lambda)\lambda}{2(\mu-\lambda)\mu} > c_d$, if 
$$
\frac{2\lambda}{2\mu-\lambda} > \frac{\lambda}{\mu-\lambda} - \lambda c_a,
$$
the equilibrium is $(0, 0)$, and if
$$
\frac{2\lambda}{2\mu-\lambda} < \frac{\lambda}{\mu-\lambda} - \lambda c_a,
$$
the equilibrium is $(1, 0)$.

For $\lambda < \mu$ and $\frac{(2\mu-\lambda)\lambda}{2(\mu-\lambda)\mu} \le c_d$, if 
$$
\frac{2\lambda}{2\mu-\lambda} > \frac{(2-d)\lambda}{2\mu-(2-d)\lambda} + \frac{d\lambda}{2\mu-d\lambda} - \lambda c_a,
$$
the equilibrium is $(0, 0)$, and if 
$$
\frac{2\lambda}{2\mu-\lambda} < \frac{(2-d)\lambda}{2\mu-(2-d)\lambda} + \frac{d\lambda}{2\mu-d\lambda} - \lambda c_a,
$$
the equilibrium is $(1, \hat{d})$.

Finally for $\lambda \ge \mu$, if
$$
\frac{2\lambda}{2\mu-\lambda} > \frac{(2-d)\lambda}{2\mu-(2-d)\lambda} + \frac{d\lambda}{2\mu-d\lambda} - \lambda c_a,
$$
the equilibrium is $(0, 0)$, and if
$$
\frac{2\lambda}{2\mu-\lambda} < \frac{(2-d)\lambda}{2\mu-(2-d)\lambda} + \frac{d\lambda}{2\mu-d\lambda} - \lambda c_a,
$$
the equilibrium is $(1, \hat{d})$.

\bibliographystyle{IEEEtran}
\bibliography{Bibliography}

\end{document}